# From the Intergalactic to the Interstellar Scales – EQUALS: a High-resolution Legacy Survey of Gas in the Distant Universe Using ESPRESSO

Trystyn Berg[1]
Valentina D'Odorico[2,3]
Elisa Boera[3]
Giorgio Calderone[2]
Rodrigo Cuellar[4]
Guido Cupani[2,3]
Stefano Cristiani[2,3]
Simona Di Stefano[5,2]
Andrea Grazian[6]
Francesco Guarneri[7,2]
Vid Iršič[8]
Sebastian Lopez[4]
Dinko Milaković[2,3]
Pasquier Noterdaeme[9]
Luca Pasquini[10]
Matteo Viel[11,2,3]
Louise Welsh[12,2]

[1] Camosun College, Victoria, Canada
[2] INAF–Trieste Astronomical Observatory, Italy
[3] Institute for the Fundamental Physics of the Universe (IFPU), Trieste, Italy
[4] University of Chile, Santiago, Chile
[5] University of Trieste, Italy
[6] INAF–Padua Astronomical Observatory, Italy
[7] Hamburg University, Germany
[8] University of Hertfordshire, Hatfield, UK
[9] IAP, Paris, France
[10] ESO
[11] SISSA, Trieste, Italy
[12] Durham University, UK

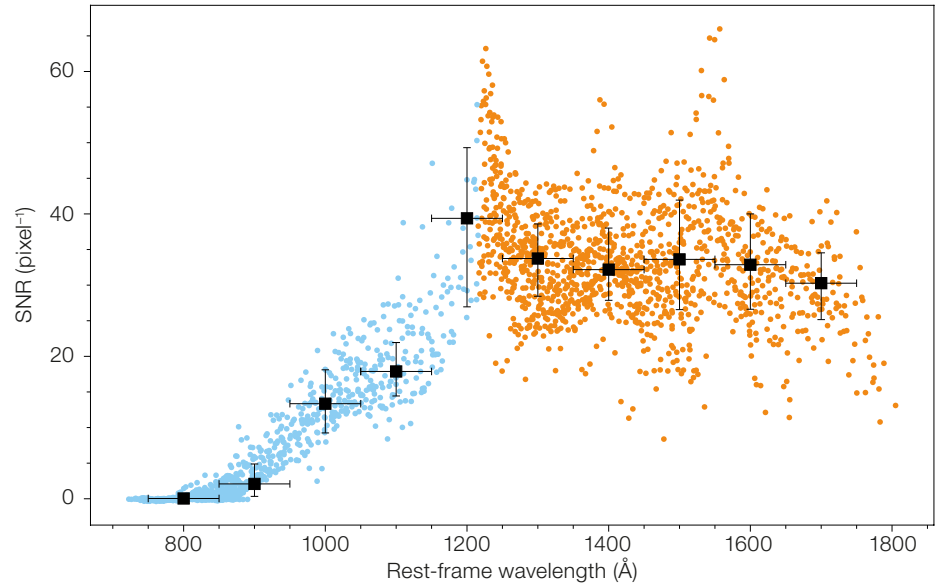

Figure 1. The range in measured SNR (per 1 km s$^{-1}$ pixel) for all 23 EQUALS spectra. The SNR is measured near regions of pure continuum and is shown as a function of the rest-frame wavelength of the respective quasar. The points are colour-coded according to whether they are blueward (light blue) or redward (dark orange) of the quasars' Lyα emission line. Black squares represent the median SNR within 100 Å bins, with the vertical error bars representing the interquartile range of SNR of each bin.

Understanding how the Universe evolved from diffuse primordial gas into the rich cosmic web we observe today is one of the great challenges of modern astrophysics. Quasar absorption lines — the imprints left by intervening gas on the light from distant quasars — provide key diagnostics of many aspects of this investigation, ranging from fundamental physics to cosmology and galaxy formation. The unprecedented combination of extremely precise wavelength calibration, high spectral resolution and high sensitivity of the Echelle SPectrograph for Rocky Exoplanet and Stable Spectroscopic Observations (ESPRESSO) has finally enabled observations that will further constrain both state-of-the-art cosmological simulations of galaxy evolution and theoretical stellar nucleosynthetic yields. In this article, we present the ESPRESSO Quasar Absorption Line Survey (EQUALS), an ESO Large Programme, designed to tackle several outstanding questions from constraining the properties of dark matter at the smallest scales probed by the Lyman-alpha forest to determining the temperature of the intergalactic medium at $z \sim 4$ and precisely quantifying the chemical contributions of stellar populations in the early Universe. EQUALS will provide a legacy sample of deep spectra to showcase ESPRESSO capabilities to the quasar absorption line community whilst providing epoch measurements for the key science goals of upcoming spectroscopic instrumentation on the next generations of telescopes.

## Survey description and data quality

Quasar sightlines are ideal for studying the gaseous reservoirs of galaxies and probing cosmology and fundamental physics. Large samples of high-redshift quasar spectra observed at high resolution ($R \sim 45\,000$) and good signal-to-noise ratio (SNR) have demonstrated their immediate and legacy values for the progress of several scientific topics (for example, Bergeron et al., 2004). Much remains to be done below the ~10 km s$^{-1}$ resolution limit imposed by the previous generation of high-resolution spectrographs. Fortunately, the Echelle SPectrograph for Rocky Planet and Stable Spectroscopic Observations (ESPRESSO; Pepe et al., 2021) can reach a much higher resolution (~2 km s$^{-1}$), enabling extragalactic science at smaller scales.

The ESPRESSO Quasar Absorption Line Survey (EQUALS) is an ESO Large Programme that was awarded 186 hours of ESPRESSO time on ESO's Very Large Telescope during Periods P112 and P113 to observe 23 bright, $z \sim 4$ quasars from the QUasars as BRight beacons for Cosmology in the Southern hemisphere (QUBRICS) survey (Guarneri et al., 2022 and references therein), quadrupling the largest homogenous sample of quasars observed with ESPRESSO (Murphy et al., 2022). The quasar sample of EQUALS was selected solely based on the brightest $z \sim 4$ QUBRICS quasars observable from the Paranal Observatory. As such, the sample is blind to selecting targets with known absorbers or previous observations. With the very high spectral resolution of the HR 4 × 2 binning mode, ESPRESSO provides an unprecedented spectral resolving power of $R \sim 140\,000$ for faint targets such as high-redshift quasars within observing times similar to those of the Ultraviolet and Visual Echelle Spectrograph (UVES) at $R \sim 40\,000$ (Berg et al., 2022).

Observations were taken in hour-long observing blocks with an additional wavelength calibration frame using the Fabry–Pérot interferometer to ensure a precise wavelength calibration at the time of observations. The single frames were reduced following Version 3.1.0 of the ESPRESSO ESO-reflex pipeline. The observations for each quasar were combined using the order-by-order spectra produced by the pipeline using custom procedures (Cupani et al., 2020). Figure 1 shows the SNR range (per $\approx 1$ km s$^{-1}$ pixel) for all 23 spectra, meeting our goal of $\sim 20$ pixel$^{-1}$ in the Lyman-alpha (Ly$\alpha$) forest and $\sim 40$ pixel$^{-1}$ redward of the Ly$\alpha$ emission of the quasar. The quality of the observations is demonstrated in Figure 2; compared to UVES, EQUALS data resolve much narrower and weaker components. EQUALS data products will be publicly released on the Canadian Advanced Network For Astronomy Research (CANFAR) and ESO's Science Archive Facility after the data reduction has been finalised.

Below we outline the science goals of EQUALS and the novel contributions it will make to probing the gaseous reservoirs of the Universe at unprecedented velocity scales.

### Probing the physics of the Ly$\alpha$ forest

The Ly$\alpha$ forest consists of the absorption lines produced by intervening neutral hydrogen along the lines of sight to distant quasars where 80–90% of the baryons at $z > 2$ reside. These intergalactic structures trace the underlying filamentary structure composed of dark matter (DM), making the Ly$\alpha$ forest a unique cosmological tool at scales and redshifts which currently cannot be probed by any other observable. The growth of cosmic structures is linked to the one-dimensional flux power spectrum in the Ly$\alpha$ forest. Spanning scales from $k < 0.01$ s km$^{-1}$ to $k \sim 0.1$ s km$^{-1}$, the flux power spectrum has been used to infer cosmological parameters (Baur et al., 2015; Chabanier et al., 2019; Palanque-Delabrouille et al., 2020), constrain DM flavours (Viel et al., 2013; Iršič et al., 2017a; Murgia et al., 2018), and measure the temperature of the intergalactic medium (IGM; Walther et al., 2018; Gaikwad et al., 2021).

The sample size of EQUALS was tuned to precisely measure the fast-declining signal at the smaller spatial scales ($k > 0.1$ s km$^{-1}$) in the Ly$\alpha$ forest, scales which are sensitive to the physics of DM and separating temperature from line broadening effects within the IGM (Boera et al., 2019). ESPRESSO's excellent resolving power (i.e. probing 10 times larger $k$) provides a factor of 20 improvement in the measurement uncertainty of the IGM temperature. Measuring the Ly$\alpha$ forest power spectrum at these smallest scales enables an improved measurement of the IGM temperature and a stronger constraint on the DM particle mass or alternative DM models (Figure 3; Viel et al., 2013; Murgia et al., 2018). Based on recent studies (Iršič et al., 2017b; Rogers et al., 2021; Karaçayli et al., 2022), we expect the improved statistical uncertainties from EQUALS will provide a factor of 5–15 gain in DM mass constraints.

### Constraining models of the baryon cycle

The physical properties of the circumgalactic medium (CGM) — including temperature, density, kinematics and ionisation state — offer vital clues about galaxy evolution. Since Cosmic Noon ($z \sim 2$–3), when star formation peaks (Madau & Dickinson, 2014), CGM studies using

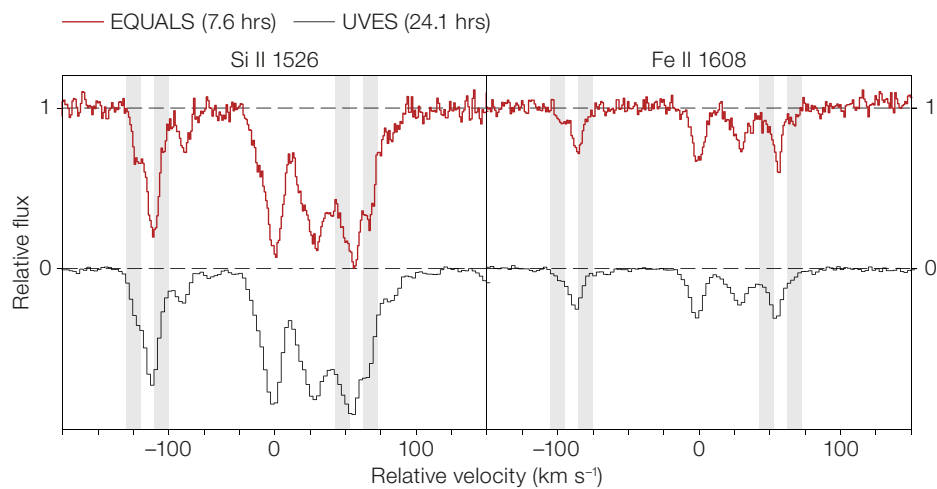

Figure 2. The normalised EQUALS spectrum of two absorption lines (Si II 1526 Å and Fe II 1608 Å) from a single absorber (red line) towards the quasar J1621-0042. The EQUALS exposure time is 7.6 hours of on-target integration, providing a SNR of $\sim 30$ and $\sim 50$ per resolution element for the Si II and Fe II lines (respectively). For comparison, the same transition lines as observed by UVES (SNR $\sim 150$ per resolution element) are shown in black. The relative fluxes of the UVES lines are shown on the same scale but are shifted vertically for visual clarity. There are many additional features that can be resolved by the extra spectral resolution of ESPRESSO, such as the four features highlighted by the light grey vertical bars.

Mg II and Fe II lines show this gas traces outflows driven by star-forming activity (Steidel et al., 2002; Churchill et al., 2020). Hydrodynamical simulations (Hummels et al., 2019; Peeples et al., 2019) predict the spatial extent and thermal properties of the CGM, both of which help distinguish inflowing IGM gas from outflowing, enriched interstellar medium ejecta.

Direct measurements of CGM gas temperature — key to distinguishing between cool, metal-poor inflows and warm, metal-rich outflows — are scarce, constrained by the need for high SNR and resolving power ($R > 100\,000$) to determine the incidence of weak Mg II absorbers (equivalent widths $< 300$ mÅ) and decouple thermal broadening from turbulence (for example, Carswell et al., 2012; Noterdaeme et al., 2021). To date, temperature measurements exist for only a small sample of strong absorbers (with equivalent widths $> 300$ mÅ). With ESPRESSO's resolution and wavelength coverage, the EQUALS sample enables simultaneous detection of metal lines from elements with large atomic mass differences (≥15 amu), thereby separating thermal and turbulent line broadening. EQUALS is expected to double the number of Mg II absorbers with reliable gas temperature estimates (Noterdaeme et al., 2021) for constraining CGM feedback models (Hummels et al., 2019; Peeples et al., 2019).

Chemical enrichment of the Universe

Intervening quasar absorption line systems are key to understanding the chemical enrichment of the Universe, from profiling the stellar population within galaxies (for example, Prochaska et al., 2003; Berg et al., 2015; Skúladóttir et al., 2018; Saccardi et al., 2023) to understanding the mechanisms for distributing metals into the IGM (Ferrara et al., 2000; Madau et al., 2001).

Very metal-poor absorbers (≥100 times more Fe-poor than the Sun; Cooke et al., 2015, Welsh et al., 2019) are excellent test cases to place meaningful constraints on the elusive properties of the first generations of stars (i.e. Population III). The initial mass function, the number of stars formed and the supernova mechanism of these Population III stars remain unknown. The isotopic ratios of $^{12}C/^{13}C$ and $^{24}Mg/^{25}Mg$ help discriminate between stellar progenitor masses and stellar populations (Heger & Woosley, 2010). $^{12}C/^{13}C$ is particularly powerful in the low-metallicity regime as it can provide observational evidence of the Population III initial mass function (Welsh et al., 2020). However, only about 20 metal-poor systems have precise abundances, and measuring isotopic ratios is challenging and relies on very small wavelengths shifts that can only be constrained at high spectral resolving powers ($R > 100\,000$) and with exquisite wavelength calibration (Welsh et al., 2020; Noterdaeme et al., 2021; Milaković et al., 2024).

Using a carefully crafted instrumental profile model, the EQUALS spectra will allow us to search for the ~3 km s$^{-1}$ shift in $^{12}C/^{13}C$ from C II lines, which can reveal the contribution from low-mass Population II and Population III stars (≤15 M$_\odot$) at 3σ significance (Welsh et al., 2020). Based on our current analysis of the data, we have about eight high-column-density systems suitable for this analysis.

To study metal enrichment in the low-density IGM, we will complement statistical methods like the pixel optical depth technique (Aguirre et al., 2002; Schaye et al., 2003) with direct detections of weak metal lines (for example, O VI, C IV) associated with low-column-density H I (log N(H I) ≤ 14–14.5). This method offers a more reliable handle on the amount of line blending and enables measurement of the enriched gas filling factor through direct H I–C IV associations (for example, Ellison et al., 2000). EQUALS will reach C IV column densities log N(C IV) ~ 11.5 — matching the deepest high-resolution efforts to date (D'Odorico et al., 2016). Within the EQUALS data, we can compute the two-point correlation function of the normalised flux and estimate the enrichment of the IGM from the signal due to the C IV doublet and to other IGM metal lines (for example, Si IV, Mg II, Fe II, Al III; Karacayli et al., 2022; Tie et al., 2022).

The legacy of EQUALS

EQUALS aims to answer today's questions and lay the groundwork for future discoveries. The high-resolution, high-SNR spectra collected from the largest homogenous sample of quasars observed with ESPRESSO will serve as a benchmark dataset for the astronomical community, enabling a wide range of future investigations. These spectra will be instrumental in refining measurements of the fine-structure constant (for example, Murphy et al., 2022; Webb et al., 2025), tracking the redshift drift over cosmic time (a key science case for the ArmazoNes high Dispersion Echelle Spectrograph [ANDES] at ESO's Extremely Large Telescope; Liske et al., 2008; Trost et al., 2025), and probing Big Bang nucleosynthesis through deuterium abundance studies (Cooke et al., 2018). ESPRESSO's exceptional resolution enables precise continuum placement and separation of Lyα forest lines from deuterium — critical steps in such analyses.

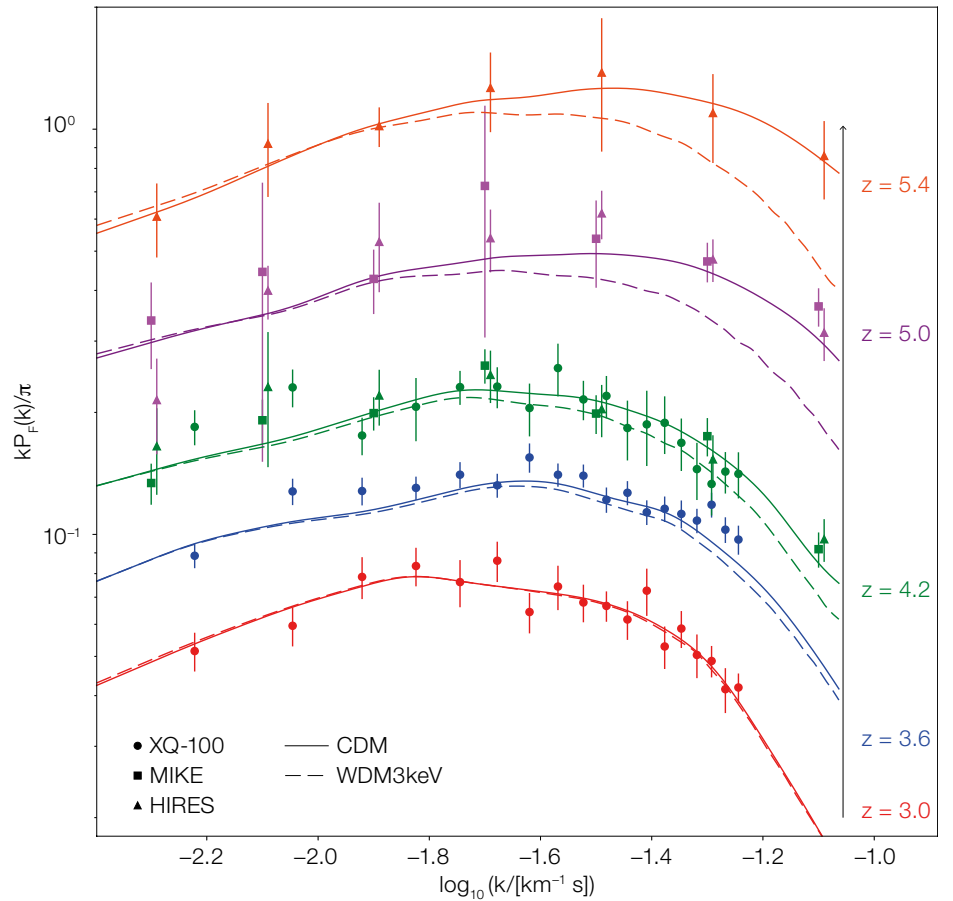

Figure 3. Constraining dark matter flavour in the Lyα power spectrum. The Lyα forest power spectrum (P(k)) as a function of scale (log k) at different redshifts (different colours). The data points represent observations obtained with different spectrographs. The solid and dashed curves denote the expectations for two different models: cold DM and 3 keV warm DM, respectively. EQUALS will probe scales of log(k/[s km$^{-1}$]) ≥ –1, where different flavours of DM have the strongest effect on the power spectrum, and the redshift range 3 ≤ z ≤ 4.3 for which these spatial scales have yet to be investigated consistently.

Just as the UVES Large Programme Cosmic Evolution of the IGM (Bergeron et al., 2004) revolutionised quasar absorption line studies and led to over a hundred publications, EQUALS will open new frontiers and inspire new questions and methodologies for years to come.

Acknowledgements

Based on observations collected at the European Southern Observatory under ESO programme 112.25NR. This research used the Canadian Advanced Network For Astronomy Research (CANFAR) operated in partnership by the Canadian Astronomy Data Centre and The Digital Research Alliance of Canada with support from the National Research Council of Canada, the Canadian Space Agency, CANARIE and the Canadian Foundation for Innovation.

References

Aguirre, A. et al. 2002, ApJ, 576, 1
Baur, J. et al. 2015, arXiv:1512.08482
Berg, T. A. M. et al. 2015, MNRAS, 452, 4326
Berg, T. A. M. et al. 2022, A&A, 662, A35
Bergeron, J. et al. 2004, The Messenger, 118, 40
Boera, E. et al. 2019, ApJ, 872, 101
Carswell, R. F. et al. 2012, MNRAS, 422, 1700
Chabanier, S. et al. 2019, JCAP, 7, 17
Churchill, C. W. et al. 2020, ApJ, 904, 28
Cooke, R. J. et al. 2015, ApJ, 800, 12
Cooke, R. J. et al. 2018, ApJ, 855, 102
Cupani, G. et al. 2020, Proc. SPIE, 11452, 114521U
D'Odorico, V. et al. 2016, MNRAS, 463, 2690
Ellison, S. L. et al. 2000, AJ, 120, 1175
Ferrara, A. et al. 2000, MNRAS, 319, 539
Gaikwad, P. et al. 2021, MNRAS, 506, 4389
Guarneri, F. et al. 2022, MNRAS, 517, 2436
Heger, A. & Woosley, S. E. 2010, ApJ, 724, 341
Hummels, C. B. et al. 2019, ApJ, 882, 156
Iršič, V. et al. 2017a, Phys Rev D, 96, 023522
Iršič, V. et al. 2017b, MNRAS, 466, 4332
Karaçaylı, N. G. et al. 2022, MNRAS, 509, 2842
Liske, J. et al. 2008, MNRAS, 386, 1192
Madau, P. et al. 2001, ApJ, 555, 92
Madau, P. & Dickinson, M. 2014, ARA&A, 52, 415
Milaković, D. et al. 2024, MNRAS, 534, 12
Murgia, R. et al. 2018, Phys Rev D, 98, 083540
Murphy, M. T. et al. 2022, A&A, 658, A123
Noterdaeme, P. et al. 2021, A&A, 651, A78
Palanque-Delabrouille, N. et al. 2020, JCAP, 4, 38
Peeples, M. S. et al. 2019, ApJ, 873, 129
Pepe, F. et al. 2021, A&A, 645, A96
Prochaska, J. X. et al. 2003, ApJS, 147, 227
Rogers, K. K. et al. 2022, Phys Rev L, 128, 171301
Saccardi, A. et al. 2023, ApJ, 948, 35
Schaye, J. et al. 2003, ApJ, 596, 768
Skúladóttir, Á. et al. 2018, A&A, 615, A137
Steidel, C. C. et al. 2002, ApJ, 570, 526
Tie, S. S. et al. 2022, MNRAS, 515, 3656
Trost, A. et al. 2025, A&A, in press
Viel, M. et al. 2013, Phys Rev D, 88, 043502
Walther, M. et al. 2018, ApJ, 852, 22
Webb, J. K. et al. 2025, MNRAS, 539, L1
Welsh, L. et al. 2019, MNRAS, 487, 3363
Welsh, L. et al. 2020, MNRAS, 494, 1411

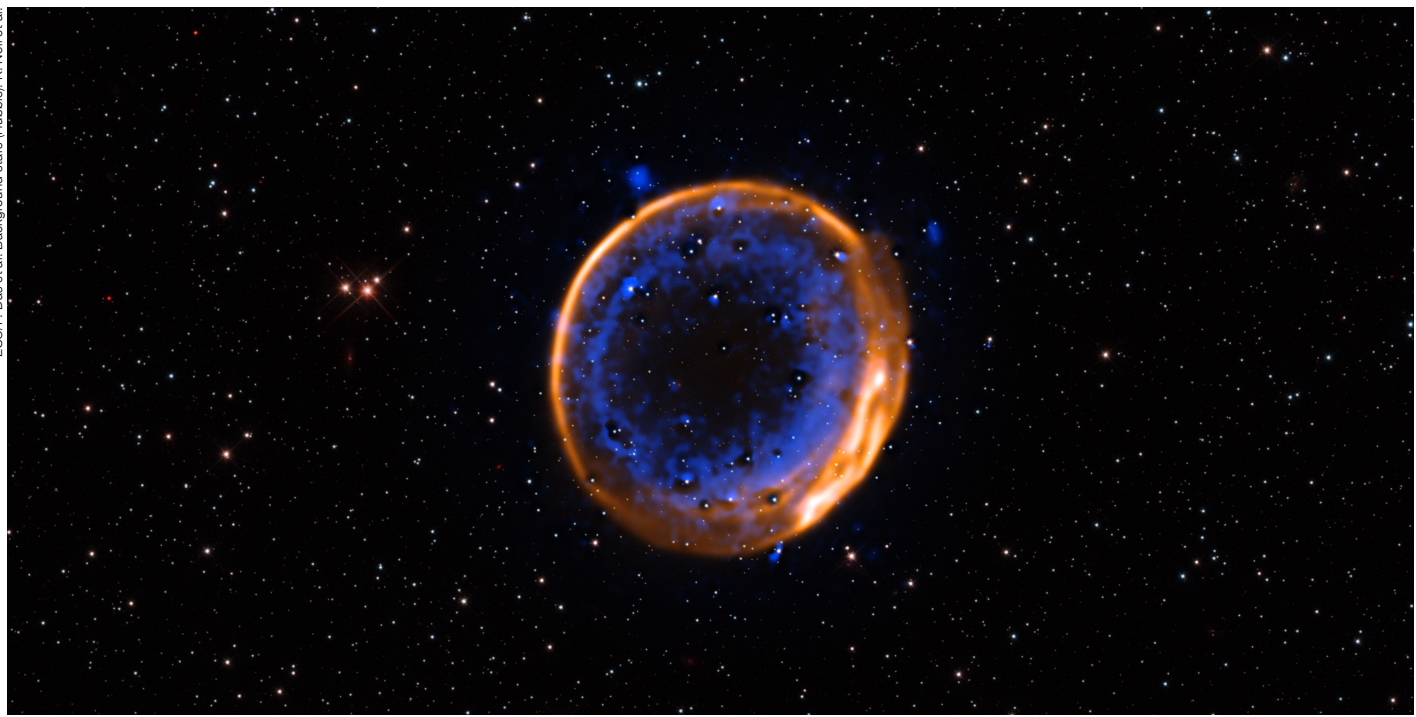

ESO/P. Das et al. Background stars (Hubble): K. Noll et al.

This image, taken with ESO's Very Large Telescope (VLT), shows the supernova remnant SNR 0509-67.5. These are the expanding remains of a star that exploded hundreds of years ago in a double-detonation — the first photographic evidence that stars can die with two blasts.